\documentclass[fleqn,twoside]{article}
\usepackage{espcrc2}

\usepackage{graphicx}
\usepackage[figuresright]{rotating}

\newcommand{\AmS}{{\protect\the\textfont2
  A\kern-.1667em\lower.5ex\hbox{M}\kern-.125emS}}

\title{
\vskip -110pt
{\large  
\mbox{} \hfill BNL-NT-04/29\\
}
\vskip 45pt
Cutoff effects in meson spectral functions
       }

\author{
T. Blum
\address{Physics Department, University of Connecticut, Storrs, 
CT 06269-3046 and RIKEN-BNL, Brookhaven National Laboratory, 
Upton, NY 11973 USA},
 P. Petreczky
\address[BNL]{Physics Department and RIKEN-BNL, Brookhaven National Laboratory,
        Upton, NY 11973 USA}
        \thanks{Goldhaber Fellow, RIKEN Fellow}                             
        }
       
\begin{document}

\begin{abstract}
We study the lattice spacing dependence
of meson spectral functions calculated
in quenched QCD with domain wall fermions
as well as clover Wilson fermions in quenched
and partially-quenched QCD. We conclude that
for lattice spacing $a \le 3$ GeV all excited
states appearing in the spectral functions 
are lattice artifacts.
\end{abstract}

\maketitle

\section{Introduction}

Temporal meson correlation functions
in Euclidean time can be related to meson
spectral functions by analytic continuation.
Furthermore using the Maximum Entropy Method
one can extract the spectral functions from
meson correlators calculated on a finite lattice
\cite{mem}. The method was applied at 
zero \cite{mem,cppacs} as well as
non-zero temperature \cite{dil,umeda02,asakawa03,datta03}.
In such an analysis one usually uses all time slices
in the correlation function not just the long
distance part. Therefore one is sensitive
to short distance physics which may be affected
by lattice artifacts. The lattice distortion of the correlators
at short distances 
translates into lattice artifacts in the spectral functions
for large energies.
So far only Wilson gauge action and Wilson (clover) fermion
action have been used to study meson spectral functions.
The lattice artifacts seen in numerical simulations 
for this combination of actions 
turns out to be quite different from what is expected in the
free theory \cite{inf}. The purpose of this paper is
to study cutoff effects of meson spectral functions 
using different gauge and fermion actions.

\section{Numerical results}

We consider the temporal correlators of meson currents
\begin{equation}
G(\tau,T)=\sum_{\vec{x}} \langle J(\vec{x},\tau) J(\vec{0},0) \rangle~,
\end{equation}
with $J=\bar q(\vec{x},\tau) \Gamma q(\vec{x},\tau)$ and
$\Gamma=\gamma_5,\gamma_{\mu}$ for
pseudo-scalar and vector channels.
For temperature $T$ the spectral function
is related to the meson correlator by the integral relation
\begin{equation}
G(\tau,T)=\int_0^{\infty} d \omega \sigma(\omega,T) 
\frac{\cosh \omega(\tau-1/(2T))}{\sinh \omega/(2T)}~.
\label{rel}
\end{equation}
\begin{figure}
\vspace*{-0.1cm}
\includegraphics[width=2.4in]{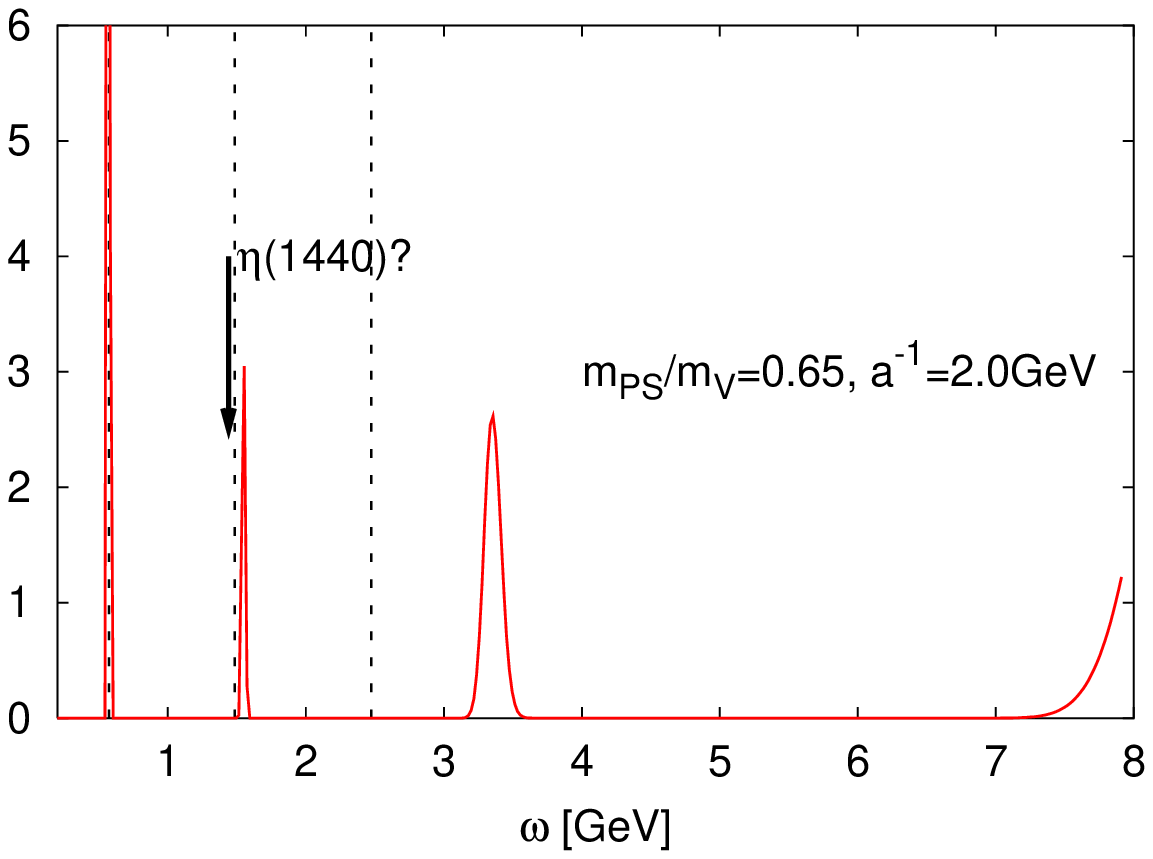}
\includegraphics[width=2.4in]{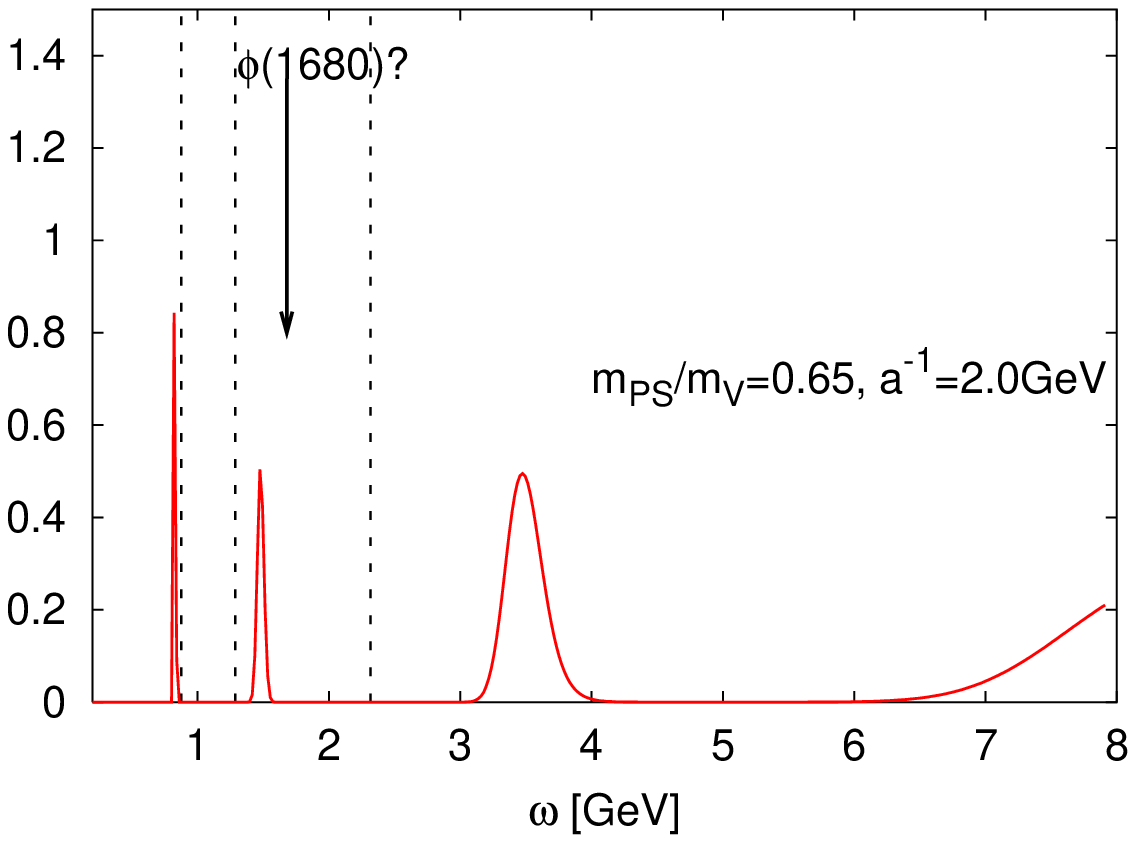}
\vspace*{-1.0cm}
\caption{The pseudo-scalar (top) and vector (bottom)
spectral functions with DWF at $a^{-1}=2.0$ GeV.
The vertical line and band show the mass of the ground
and first excited states.
}
\label{dwf2.0_fig}
\end{figure}
Although this relation is valid {\it a priori} only in the continuum,
it has been shown recently that in the limit of a non-interacting 
theory it holds on the lattice \cite{inf} as well.
 
We use quenched domain wall fermon correlators
on DBW2 gauge configurations from the RBC collaboration.
Furthermore, we calculate
tree-level clover fermion correlators on quenched 1-loop
tadpole-improved Symanzik lattices
and on partially-quenched 2+1 flavor
Asqtad fermion lattices from the MILC collaboration.
 
To extract the
spectral function we use the Maximum Entropy Method.
We also parametrize the spectral functions as a sum of
delta functions $\sigma(\omega)=\sum_i F_i \delta(\omega^2-m_i^2)$
and determine the parameters $F_i$ and $m_i$ using
constrained curve fitting techniques\cite{lepage} which allows for
better systematic control of the results.

\begin{figure}[h]
\vspace*{-0.4cm}
\includegraphics[width=2.4in]{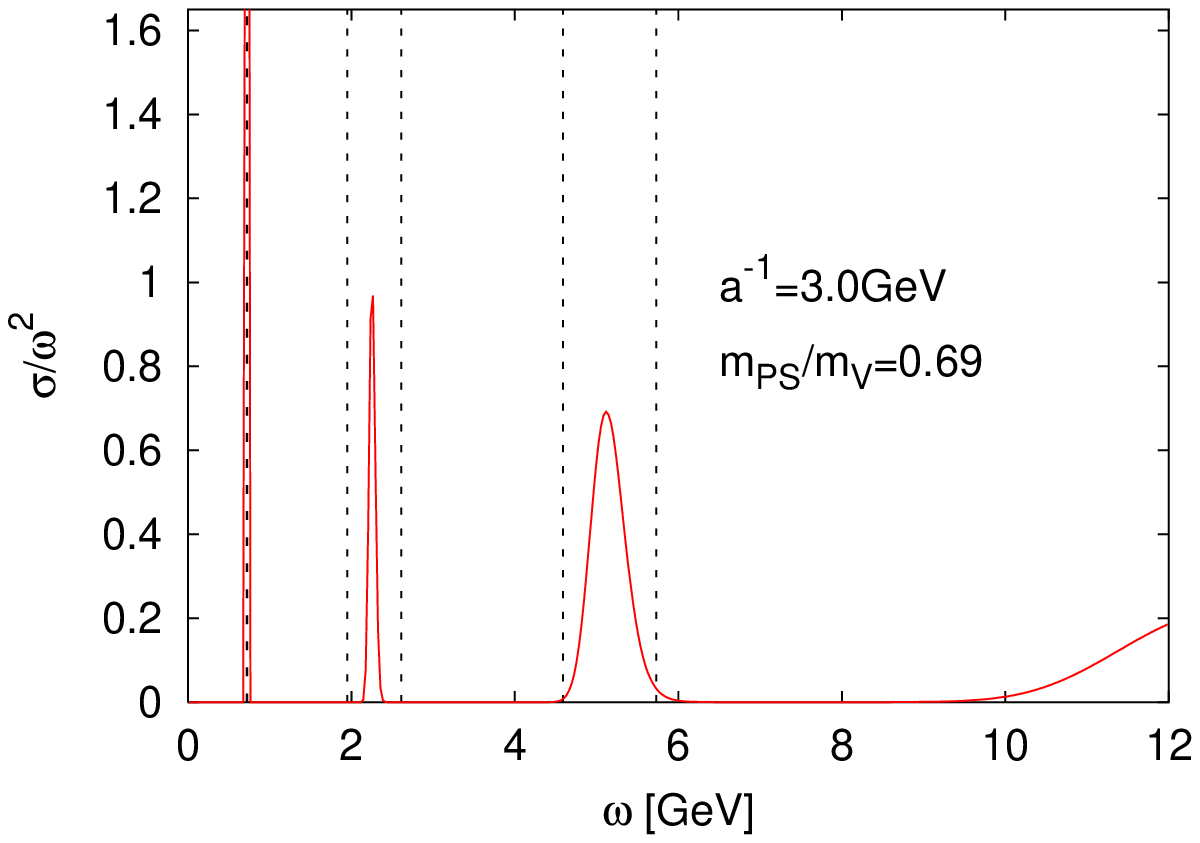}
\includegraphics[width=2.4in]{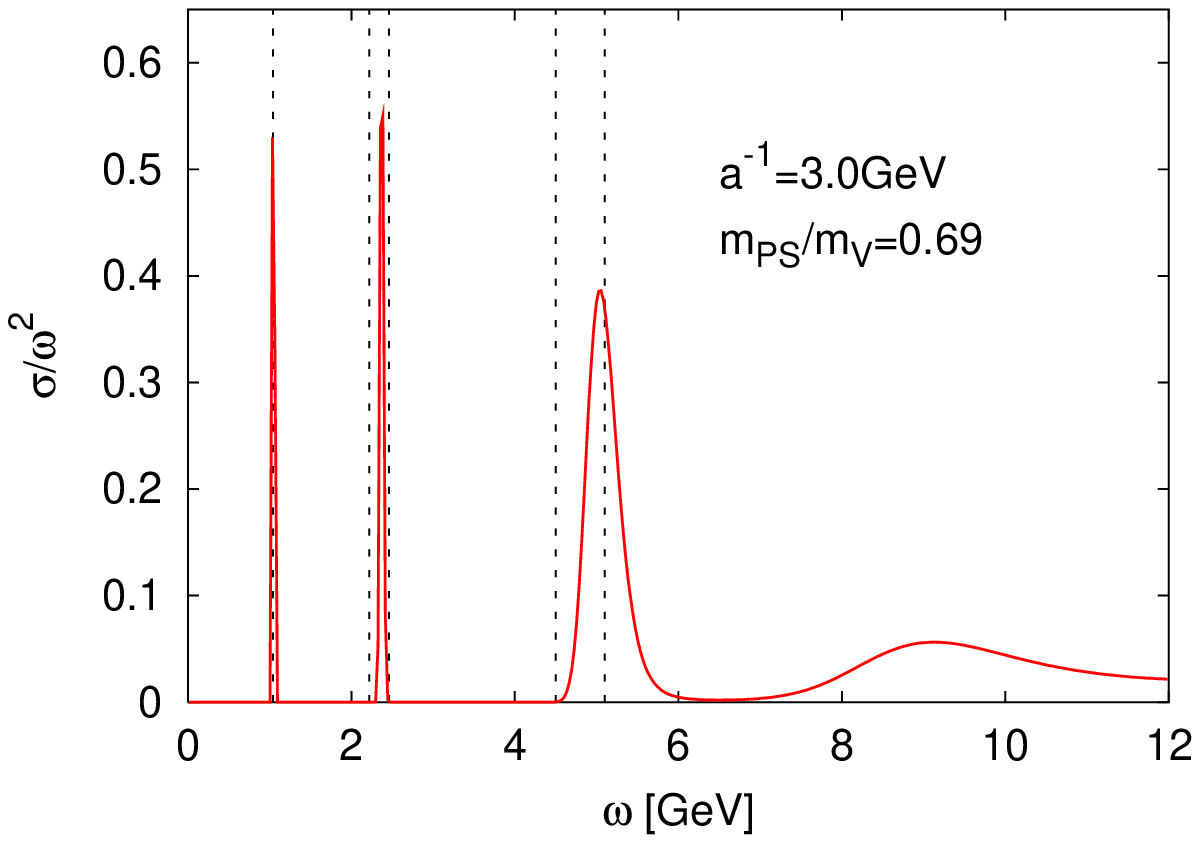}
\vspace*{-1.0cm}
\caption{The pseudo-scalar (top) and vector (bottom)
spectral functions with DWF at $a^{-1}=3.0$ GeV. The vertical
bands indicate the error on the peak location as determined from a constrained fit}
\label{dwf3.0_fig}
\end{figure}

We start the discussion of our numerical results with the case
of domain wall fermions. Here we performed calculations at three
different lattice spacings, $a^{-1}=1.3,~ 2.0,~3.0$ GeV and valance 
quark mass around the strange quark mass. For the
coarsest lattice we see a two peak structure in the spectral
functions with the second peak 
located around 2GeV. The positions of
the peaks are reasonably reproduced using  a two exponential fit.
In Fig. \ref{dwf2.0_fig} we show our results for finer lattices,
$a^{-1}=2.0$ GeV for the pseudo-scalar and vector spectral  functions.
We see a three peak structure, with the first excited state being close
to the  position of the physical excited states 
of the $\eta$ and $\phi$ mesons. The third peak is likely to be
a lattice artifact. We also performed a simple two exponential
fit of the corresponding meson correlators. The results of the 
fit are shown as the vertical line and as a band. The ground state
mass from the fit (statisitcal error $\sim 1$\% not shown) is in 
reasonable agreement with the MEM analysis
while for first excited state there is larger statistical uncertainty 
indicated by the band. To see whether or not the
second and third peak correspond to physical resonances or 
lattice artifacts, we calculated  the spectral functions at
$a^{-1}=3.0$ GeV. The results are shown in Fig. \ref{dwf3.0_fig}
which shows that the ground state peak did not shift 
(apart from a small shift due to slightly larger
quark mass) while the second and the third peaks have moved 
roughly as $a^{-1}$. This indicates that these peaks are lattice 
artifacts. Similar results were found for the Wilson action
\cite{mem,cppacs}. In Ref. \cite{cppacs} they were interpreted as
bound states of Wilson doublers. For DWF, however,
such an interpretation is more difficuilt to apply.
For this lattice spacing we have also performed a constrained 
fit with three exponentials. The ground state mass shown 
as the vertical line in Fig. \ref{dwf3.0_fig} was obtained from 
a simple uncostrained fit which was then used as a prior 
for the constrained fit. The result for the second
and third peak from the constrained fit is shown as bands in 
Fig. \ref{dwf3.0_fig}. Reasonable agreement between the MEM
and constrained curve fits is found.

\begin{figure}
\vspace*{-0.2cm}
\includegraphics[width=2.4in]{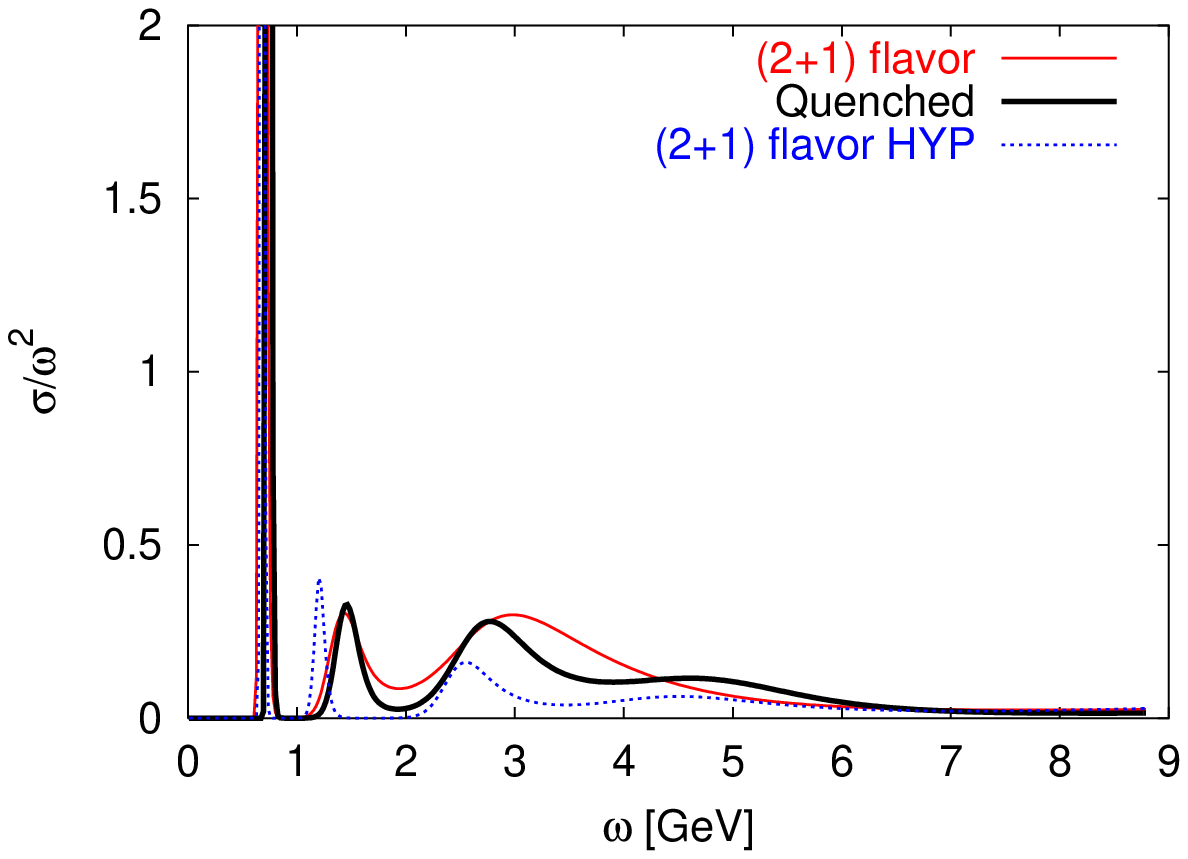}
\includegraphics[width=2.4in]{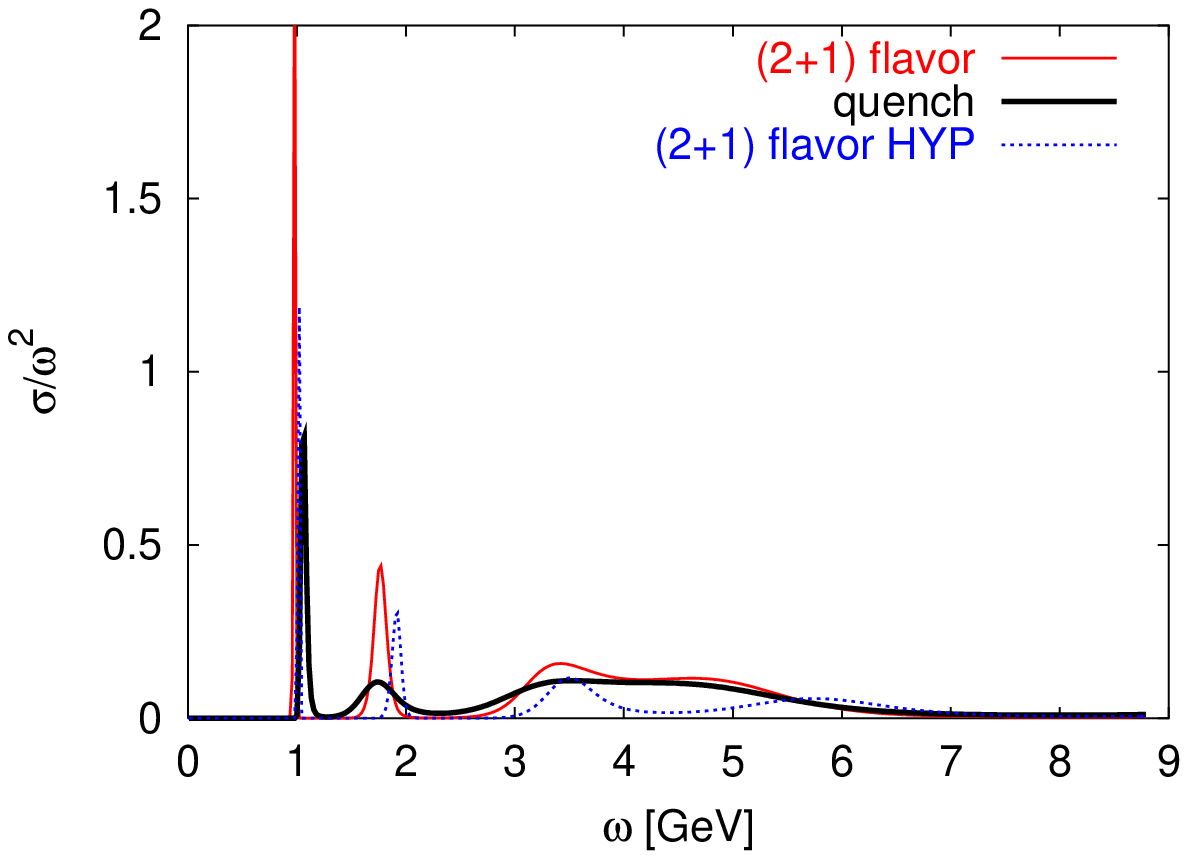}
\vspace*{-1.0cm}
\caption{Pseudo-scalar (top) and vector (bottom) spectral functions 
for clover action in quenched and partially-quenched QCD.}
\label{wilson_fig}
\end{figure}

One may wonder whether the second and third peaks 
are artifacts of quenching. Absent dynamical quarks,
all hadrons are absolutely stable (i.e. have zero width) 
no matter how large their mass
is. If dynamical quarks are present, hadrons may decay, 
and the higher
the mass the larger the width. Therefore higher lying resonances
merge into a continuum. To check this we performed calculations 
in quenched and partially-quenched QCD
using the clover action.
More precisely we used  gauge configurations with Symanzik action
generated by the MILC collaboration as well the
dynamical 2+1 Asqtad configurations. In both cases
the lattice spacing was $a\approx0.09fm$.
We use the tree-level value for the clover coefficient, $c_{sw}=1$, and  
$\kappa=0.14$ which corresponds to $m_{PS}/m_{V}=0.68$.
Results for the spectral functions are shown in Fig. \ref{wilson_fig}.
As one can see, the spectral functions for quenched and partially 
quenched QCD are very similar, indicating that for quark masses 
around the strange quark mass effects due to quenching are quite
small. 
In addition we have also calculated meson spectral functions 
wit HYP smeared clover action in $2+1$ partially-quenched QCD,
i.e. we have applied HYP smearing on the dynamical configuations 
before calculating the meson correlators. 
We have performed calculations at several quark masses.
From extrapolation of the pseudo-scalar mass squared to zero we determine 
the critical hopping parameter to be $\kappa_c=0.12718(2)$ (statistical error only).
The purpose of
calculation with HYP smearing was to see whether the smoothing of the gauge 
fields removes the artificial peaks, and the corresponding results
are shown in Fig. \ref{wilson_fig}. As one can see from the
figure, the lattice artifacts are reduced but not removed
completely by HYP smearing. However, as a by-product we observe 
that number of exceptional configurations as well 
as the number of CG in the fermions matrix inversions are greatly reduced.
In fact we did not observe any exceptional configurations in a 
sample of 59 configurations 
down to quark masses $\sim m_s/8$ ($m_s$ is the strange quark mass).

\section*{Acknowledgment}
\noindent
This work has been authored under the contract 
DE-AC02-98CH10886 with the U.S. Department of energy.
TB was partly supported by the LDRD program of Brookhaven
National Laboratory, Project No. 04-041. We
thank the RBC collaboration for the DWF correlators.
The program for the MEM analysis was provided by Ines Wetzorke
to whom we are very grateful.
The calculations were performed on the IBM-SP supercomputer 
at NERSC using gauge configurations from the MILC collaboration
and the MILC code for which we are grateful.


\begin{thebibliography}{99}
\bibitem{mem}
Y. Nakahara {\it et al.}, Phys. Rev. D {\bf 60} (1999) 091503;
M. Asakawa {\it et al.}, Prog. Part. Nucl. Phys. {\bf 46} (2001) 459
\bibitem{cppacs}
Yamazaki, {\it et al.}., Phys. Rev. D65 (2002) 014501
\bibitem{dil} 
F. Karsch, {\it et al.}, Phys. Lett. {\bf B530} (2002) 147
\bibitem{umeda02} T. Umeda {\it et al.}, hep-lat/0211003 
\bibitem{asakawa03}
M. Asakawa and T. Hatsuda, Phys.Rev.Lett. 92 (2004) 012001
\bibitem{datta03}
S. Datta, {\it et al.},  Phys.Rev. D69 (2004) 094507
\bibitem{inf}
F. Karsch, {\it et al.}, Phys. Rev. D {\bf 68} (2003) 014504
\bibitem{lepage}
G.P. Lepage, {\it et al.}.,
Nucl.Phys.Proc.Suppl. 106 (2002) 12-20
\end{thebibliography}
\end{document}